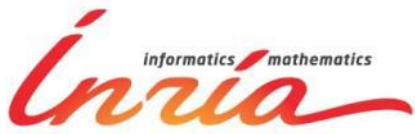

# The control over personal data: True remedy or fairy tale ?

Christphe Lazaro, Daniel Le Métayer

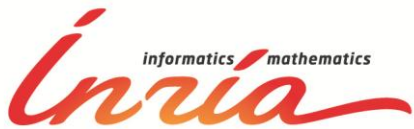

# The control over personal data : True remedy or fairy tale ?

Christophe Lazaro[1] , Daniel Le Métayer [2],

**Abstract:**  This research report undertakes an interdisciplinary review of the concept of "control" (i.e. the idea that people should have greater "control" over their data), proposing an analysis of this concept in the field of law and computer science. Despite the omnipresence of the notion of control in the EU policy documents, scholarly literature and in the press, the very meaning of this concept remains surprisingly vague and under-studied in the face of contemporary socio-technical environments and practices. Beyond the current fashionable rhetoric of empowerment of the data subject, this report attempts to reorient the scholarly debates towards a more comprehensive and refined understanding of the concept of control by questioning its legal and technical implications on data subject's agency.

**Key-words:** control, personal data, privacy, law, consent

[1] CRIDS, rue de Bruxelles 61, B-5000 Namur, Belgique– Christophe.lazaro@unamur.be

[2] Inria, Antenne La Doua, Bâtiment CEI-2. 56, Boulevard Niels Bohr. CS 52132. 69603 Villeurbanne, France – daniel.le-metayer@inria.fr

# The control over personal data : True remedy or fairy tale ?

**Résumé :** Ce rapport décrit les résultats d'une recherche interdisciplinaire sur la notion de contrôle sur les données personnelles. L'idée selon laquelle les individus devraient garder le contrôle sur leurs données personnelles est désormais prévalente dans de nombreux discours sur la protection de la vie privée aussi bien que dans des documents officiels d'UE. Pourtant ce concept de contrôle reste vague et assez peu analysé en tant que tel. Ce rapport tente de combler ce manque en questionnant la notion de contrôle d'un point de vue juridique et informatique en étudiant notamment ses implications sur la notion d'"agency" du sujet.

**Mots clés :** contrôle, données personnelles, vie privée, droit, consentement

# The control over personal data : True remedy or fairy tale?

**Christophe Lazaro, Daniel Le Métayer**
**Inria, University of Lyon**

**Introduction**

As personal data processing technologies change and create new possibilities to track and trace individuals, politicians and lawyers struggle to deal with their implications on informational privacy and data protection. Many claim that these problems can be tackled with improved statutory drafting techniques and call for legislation that would give individuals greater *control* over the processing of their data. More than ever the notion of control dominates the contemporary conceptual and normative landscape of data protection and privacy.

Until now it has been often advocated as the key solution to face the problems raised by personal data processing technologies. Indeed, control is considered as a precious means of empowerment of the "digital self": it is deemed to foster autonomy and ability to manage information about oneself and to have some limited dominion over the way he/she is viewed by society. For this reason, the notion of control is often mentioned, either explicitly or implicitly, as a core element of data protection policies. In the recent EU Proposal for a general data protection regulation, a set of legal instruments are potentially designed to put the data subject in control of his data such as the explicit consent requirement, the withdrawal of consent, the right to object, the right to be forgotten, the right to data portability, etc.

Despite the omnipresence of the notion of control in the EU policy documents, scholarly literature and in the press, the very meaning of the notion of control as well as its normative implications remains surprisingly vague and under-studied in the field of data protection and privacy studies.

From a fundamental rights perspective, control conceived as a set of "micro-rights" should undoubtedly be considered as a central element of any empowerment policy in the field of privacy and data protection. However, in the face of recent technological developments and emergence of new social practices which seem to undermine the very capacity or even the will of individuals to "self-manage" their informational privacy, this apparently simple and familiar notion becomes very ambiguous. What does it really mean to be in control of one's data in the context of contemporary socio-technical environments and practices?[3] What are the characteristics, the purposes and potential limits of such control? How to guarantee to individuals an effective control over their own data? Is legislation on data protection an appropriate instrument to ensure individual control?

The ambiguities raised by the notion of control are best illustrated by the famous "privacy paradox" that has been largely commented in the literature. This paradox suggests that while internet users contend that they are concerned by their privacy and complain that they are not properly informed about what happens and how their data are exploited, yet they are often willing to give away very detailed personal information either in exchange for a bargain or even for nothing on their social networks' profiles.

---
[3] See online the Symposium dedicated to "The Privacy Paradox. Privacy and Its Conflicting Values", *Stan. L. Rev. Online*, February 02, 2012 - April 12, 2012, http://www.stanfordlawreview.org/online/privacy-paradox.

Interestingly, the tension constitutive of this paradox opposes two different ways of grasping the issue of control[4]. The first one is structural/objective and relates to the risks associated to what G. Deleuze used to call "control societies"[5]. In that respect, control refers to the notion of surveillance which can be exercised by public institutions or private companies in order to monitor, regulate and influence people's behavior. The thematic of *control as surveillance* has been largely covered in the literature these last decades and has been recently reactivated after the NSA scandal and, more largely, the emergence of a new form of "algorithmic governmentality"[6]. The second way to apprehend the notion of control is individual/subjective and relates to the multiple ways individuals interact with each other and communicate personal information, and therefore participate in the self-construction of their digital identities. In that sense, c*ontrol as agency* refers to self-determination over information about oneself and self-management of privacy.

This second way to grasp the notion of control will be the central focus of the report. If one has to acknowledge that disclosure of personal data is part of contemporary everyday life and practices of individuals, we contend that it is urgent to understand the meaning of the notion of control as well as to apprehend its pragmatic modalities. Hence, it is necessary to take the current rhetoric of control and empowerment seriously as it is advocated in EU policy documents and by the proponents of the "privacy as control" theory.

This report undertakes an interdisciplinary review of the concept of "control", proposing an analysis of this concept in the field of law and computer science. Part I explores the meanings of the concept of control as it is developed in data protection scholarship and EU law. Beyond the current fashionable rhetoric of empowerment of the data subject, this part aims at identifying and critically assessing the characteristics of the concept control and its normative consequences. Part II examines the concept of control from a technical point a view. It reviews and analyses the potentialities of various technical tools in making control more efficient. To conclude, Part III attempts to reorient data protection scholarship towards a more comprehensive and refined understanding of the concept of control. In particular, it takes up the claim that taking control seriously requires to focus more strenuously on two fundamental and intertwined issues: control of *what* and control by *whom*?

---

[4] Fuchs, O., (2011), Towards an alternative concept of privacy, *JICES*, Vol. 9, No. 4, p. 222.
[5] See the concept of "les sociétés de contrôle" forged by Deleuze, G. (1990/2003), *Pourparlers 1971-1990*, Les Editions de Minuit, pp. 229-239 ; Haggerty, K. D. & Ericson, R. V. (2000), The surveillant assemblage. In: *British Journal of Sociology*, Vol. 5, No. 4, pp. 605-622.
[6] Rouvroy, A. & Berns, Th. (2013), Gouvernementalité algorithmique et perspectives d'émancipation : le disparate comme condition d'individuation par la relation? In : *Politique des algorithmes. Les métriques du web. Réseaux*, Vol.31, No. 177, pp. 163-196. Available at: http://works.bepress.com/antoinette_rouvroy/47. See also Cheney-Lippold J. (2011), A New Algorithmic Identity. Soft Biopolitics and the Modulation of Control. In: *Theory, Culture & Society*, Vol. 28, No. 6, pp. 164-181.

### I. The concept of control in data protection and privacy scholarship

The concept of control is of paramount importance in the literature dedicated to privacy and data protection. This concept is not new and long before the internet era control over personal information has been considered as one of the dominant definitions of "privacy". The "privacy as control" theory has been constructed in reaction against the definition of privacy as a "right to be let alone" advocated by the famous attorneys Brandeis and Warren at the end of the nineteen century[7]. Conceived in this fashion, privacy is a condition of insulation deemed to guarantee freedom from interference or intrusion upon personal sphere. For many scholars, this conceptualization of privacy abusively conflates privacy with liberty and misleadingly suggests the existence of a private sphere, a "bubble" that surrounds the self and into which other individuals and organizations cannot encroach[8]. Instead, the proponents of the control theory argue that privacy has nothing to do with protecting one's space from intrusion, but is determined by the ability to control personal information[9].

In that regard, the most influential formulation of privacy is perhaps the one proposed by A. Westin (1967) who describes privacy as "the claim of individuals, groups, or institutions to determine for themselves when, how, and to what extent information about them is communicated to others"[10]. The idea that privacy is the ability of the individual to control the terms under which personal information is acquired and used has been endorsed by a broad community of scholars. C. Fried (1968) also recognizes that being able to maintain control over personal information is crucial. It allows us to create the necessary context for relationships of respect, trust, friendship and trust. For this author, "privacy is not simply an absence of information about us in the minds of others; rather it is the control we have over information about ourselves"[11]. Along the same lines, J. Rachels (1975) argues that there is a close connection between the ability to control who has access to one's information and one's ability to maintain a variety of social relationships with different types of people[12]. W. Parent (1983) also tried to provide a more detail account of the control theory that does not overlap with other familiar values such as liberty, autonomy, solitude, secrecy, etc. He defines privacy in narrow terms as the condition of not having undocumented personal information known by others; therefore recognizing the importance of choice and control about "facts that most persons in a given society choose not to reveal about themselves or facts about which a particular person is extremely sensitive and which he therefore does not choose to reveal about himself"[13].

---

[7] Brandeis L., & Warren S. (1890), The Right to Privacy. In: *Harvard L. Rev.*, Vol. 4, pp. 193-220.
[8] Bennett, C. J. (2011), In Defense of Privacy : The Concept and the Regime. In: *Surveillance & Society*, Vol. 8, No. 4, p. 488.
[9] H. T. Tavani (2000), Privacy and the Internet. In: *B.C Intell. Prop. & Tech.*, p. 2, http://www.bc.edu/bc_org/avp/law/st_org/iptf/commentary/content/2000041901.html. As H. Tavani points out the conception of privacy has evolve from one concerned with intrusion and interference to one that has been concerned with information: "[…] it must be noted that recent theories of privacy have tended to center on issues related to personal information and to the access and flow of that information, rather than on psychological concerns related to intrusion into one's personal space and interference with one's personal affairs."
[10] Westin, A. F. (1967). *Privacy and freedom*, Atheneum Press, New York.
[11] Fried, Ch. (1968), Privacy. In: *Yale L. J.*, Vol. 77, p. 482.
[12] Rachels, J. (1975), Why privacy is important. In: *Philosophy & Public Affairs*, Vol. 4, p. 326.

Nowadays, the "privacy as control" theory is more vivid than ever and has been also endorsed by more recent commentators dealing with contemporary issues raised by complex digital environments, practices and devices. Therefore, current literature focuses more on informational privacy and data protection issues than on privacy *stricto sensu*[14] and the concept of control is more than ever advocated as the key solution to face the problems raised by current personal data processing technologies[15]. Control is not only mentioned as a core element of conceptual reflections, but also as a prescriptive remedy proposed by scholars[16].

Current "privacy as control" theories emphasize the role of choice and individual self-determination over other values. In that regard, they can be described as information management theories where control is achieved through the subjective management and expression of personal preferences[17]. Accordingly, individuals are deemed to be able to determinate was it good for themselves and consequently decide to withhold or disclose more or less personal information to the others[18]. Control is then conceptualized as an individual, dynamic and flexible process whereby people can make themselves accessible to others or close themselves. As M. Birnhack puts it, privacy as control is "[…] the view that a right to privacy is the control an autonomous human being should have over his or her personal information, regarding its collection, processing and further uses, including onward transfers."[19] In this view, control takes the shape of the right of individuals to know what information about themselves is collected, to determine what information is made available to third parties, and to access and potentially correct their personal data.

Beside self-determination and self-management, informational privacy scholars have also conceptualized control and data subject's rights in terms of property. Indeed, an important part of the privacy literature has focused on property-based metaphors and descriptions to sustain the argument that a greater control over personal information could be achieved through market-oriented mechanisms based on individual ownership of personal data. According to this view privacy can be compared to a property right: "[P]rivacy can be cast as a property right. People should own information about themselves and, as owners of property, should be entitled to control what it is done with it"[20]. Or as V. Bergelson puts it "[i]n order to protect privacy, individuals must secure control over their personal information by becoming real owners."[21]

---

[13] Parent, W. A. (1983), Recent Work on the Concept of Privacy. In: *American Philosophical Quarterly*, Vol. 20, No. 4, p. 341. For Parent, personal information is "undocumented" in the sense that this information does not belong to the public record.

[14] For the difference between privacy and data protection, see Hustinx P. (2005), Data Protection in the European Union. In: *Privacy & Informatie*, pp. 62-65.

[15] Peppet, S. R. (2011), Unraveling Privacy : The Personal Prospectus and the Threat of a Full-Disclosure Future. In: *Northwestern U. Law Rev.*, Vol. 105, p. 1183: "But even a cursory review of the literature should suffice to demonstrate that control dominates as the primary solution of privacy advocates".

[16] Schwartz, P. M. (2000), Internet Privacy and the State. In: *Conn. L. Rev.*, Vol. 32, p. 820 ("The leading paradigm…conceives of privacy as a personal right to control the use of one's data"); Cohen, J. E. (2000), Examined Lives: Informational Privacy and the Subject as Object. In: *Stan. L. Rev.*, Vol. 52, p. 1379 ("Data privacy advocates seek … to guarantee individuals control over their personal data."); Calo M. R. (2011), The Boundaries of Privacy Harm. In: *Ind. L.J.*, Vol. 86, p. 1134 (describing privacy harms as "the loss of control over information about oneself or one's attributes"); Litman, J. (2000), Information Privacy/Information Property. In: *Stan. L. Rev.*, Vol. 52, p. 1286 ("[A]ctual control [of information] seems unattainable.").

[17] Solove, D. J. (2013), Privacy Self-Management and the Consent Paradox. In: *Harvard L. Rev.*, Vol. 126, pp. 1880-1903.

[18] Fuchs, Ch. (2011), *op. cit.*, p. 223.

[19] Birnhack, M. D. (2011), A Quest for A Theory of Privacy : Context and Control. In: *Jurimetrics*, Vol. 51, No. 4, p. The author draws this definition from Allan Westin's 1967 seminal article.

In this view, the control over one's personal data is directly connected to the idea of legal or beneficial ownership of them[22]. In that sense, the concept of control evokes this kind of absolute power or sovereignty over things conventionally associated with the property regime. Such a basic conception of property entails an "exclusivity axiom" which theoretically allows the owner to protect one's good from unwanted uses and grant him fully alienable rights[23]. In this conception, free alienability is considered as a quintessential aspect of any logics of propertization of personal data and controlling one's data would then legally mean being entitled to trade and exchange them on the "privacy market".

Despite being anchored in completely different legal backgrounds, both of these conceptions (control as self-determination and as property) share common theoretical assumptions about privacy that originated in liberal worldviews. Indeed, the concept of control is strongly associated with the conventional figure of the "rational and autonomous agent", capable of deliberating about personal goals, controlling the course of the events and acting under the direction of such deliberation. The sense of control that the liberal picture relies on is, first, *individualist*, in the sense that it emphasizes individual choice, self-governance, and, overall, self-direction of one's life[24]. In that regard, the concept of control seems to confer to the data subject an extraordinary kind of sovereignty. For it permits each individual to define, unilaterally and independently, his relationships with others. Moreover, in this view, privacy is conceived as the separation of the self from others and society at large. Second, it is *active*, in the sense that it stresses agency and construction of a life for oneself[25]. In this view, control over personal data control cannot be reduced to the mere exercise of one's right to be let alone. Instead it refers to individual's ability or willingness to make decisions to control the use and sharing of information through active choice. Therefore, this active choice implies, on one hand, an effective participation of the data subjects in the management of their data[26], on the other hand, a liberty to alienate their data as long as this choice and the subsequent alienation is informed and voluntary.

---

[20] Litman, J., (2000), *op. cit.*, p. 1286.
[21] Bergelson, V. (2003), It's Personal but Is It Mine?, Toward Property Rights in Personal Information. In: *U.C. Davis L. Rev.*, Vol. 37, p. 383.
[22] Baron, J. B. (2012), Property as Control : The Case of Information. In: *Michigan Telecom. & Tech. L. Rev.*, Vol. 18, 2012, p. 409 ("My argument has been that medical and other information is in at least one important way alike: it is information over which individuals seek control. It is control, I have argued, that has led to calls for the propertization of information.")
[23] Rose, C. M. (1998), Canons of Property Talk, or, Blackstone's Anxiety. In: *Yale L. J.*, Vol. 108, p. 603.
[24] Let's note that to this individualistic conception of the human agent corresponds an instrumental conception of technical artifacts. These are reduced to mere tools at the disposal and at the service of the autonomous and intentional subject. Action is conceived as the rational execution of plan and presupposes the participation of artefacts reduced to functional entities. In this model, there is a radical separation between a subjectivity which intrinsic to the human agent and an objectivity hold by the extern reality (instrument). See Thévenot, L. (2002), Which road to follow ? The moral complexity of an 'equipped' humanity. In: Law, J. & Mol, (eds.), *Complexities : Social Studies of Knowledge Practices*, Durham and London, Duke University Press, pp. 53-87.
[25] Frey, R. G. (2000), Privacy, Control and Talk of Rights. In: *Social Philosophy and Policy*, Vol. 17, No. 2, pp. 45-67.
[26] A general assumption of the control theory is that data subjects can be expected to behave as if they are performing a calculus (cost-benefit analysis) in assessing the outcomes they will receive as a result of information disclosure.

At this stage of our analysis, it should be noted that both *individualistic* and *active* dimensions of control have been subject to criticisms, which sometimes are formulated by the proponents of the control theory themselves. Space does not permit an exhaustive overview of the objections formulated against the concept of control. However, it is important to note that some scholars have developed more nuanced conceptual accounts. On one hand, recent attempts in the literature aim at overcoming strictly individualistic accounts of privacy by paying attention to the collective aspects of privacy which is conceived as a common good or a social value[27]. On the other hand, some scholars have tackled the complex issues raised by alienability and control ceases to be conceived as an absolute and exclusive power, but as a prerogative which in some instances can or need to be restricted[28].

Despite these various attempts to refine the control theory in the field of privacy and data protection law and ethics, some commentators consider that the concept of control still remains too vague and ambiguous[29]. Although we do not disagree that control is a crucial issue, we share this argument.

We believe that, when defined purely in managerial terms, the concept of control or the control theories can hardly be disentangled from other privacy theories. Indeed, the concept of privacy encompasses a myriad of definitions which all require to a certain extent some level of control from the user. In the literature, the systematic inclusion of elements of control in definitions of privacy is particularly obvious in the various attempts of classification proposed by different authors. For instance, in the taxonomy developed by D. Solove, different types of privacy are enumerated. Alongside "control over personal information", 5 other types of definitions are mentioned: (1) the right to be left alone; (2) limited access to the self; (3) secrecy; (4) personhood; and (5) intimacy[30]. It is truly not clear what specifically distinguishes control over personal information from other types of actions or interests regarding privacy. To be sure, to be able to limit access, to ensure one's right to be let alone, to secure confidentiality individuals must be able to exercise some degree of control over their personal information. If narrowly conceived in managerial terms, the concept of control seems to be an essential characteristic of any definition of privacy and loses a great deal of its potential significance. Even among restricted access theories, which are conventionally opposed in the literature to control theories[31], references to the notion of control seem unescapable. In these theories, one element of control concerns avoiding unwanted intrusion or interference by others into one's private space and, consequently, implies the limitation of other's access to the self[32] and the control of personal boundaries[33] and environment[34].

---

[27] J. E. Cohen (2001), "Privacy, Ideology, and Technology : A Response to Jeffrey Rosen", *Geo. L. J.*, Vol. 89, p. 2039: "[w]e have tried to move the concept of privacy well beyond control and individual consent, reconceptualizing it in various ways as a social good that deserves protection for reasons beyond individual welfare."

[28] Schwartz, P. M. (2004), Property, Privacy and Personal Data. In: *Harvard L. Rev.*, Vol. 117, No. 7, pp. 2055-2128.

[29] Shoemaker, D.W. (2010), Self-exposure and exposure of the self-informational privacy and the presentation of identity. In: *Ethics Inf. Technol.*, Vol. 12, p. 4. As Shoemaker points it out, defining control simply as a matter of information management does not say anything about the extent of the control required or about what specifically counts as a relevant zone of personal information that should be kept under control. See also Tavani, H. T. (1999b), KDD, data mining, and the challenge for normative privacy. In: *Ethics and Information Technology*, Vol. 1, pp. 265-273.

[30] Solove, D. (2008), *Understanding Privacy*, Harvard University Press, Cambridge M.A.

[31] Tavani, H. T. (2008), Informational privacy: concepts, theories and controversies. In: Himma, K. E. & Tavani, H. T. (eds.), *The Handbook of Information and Computer Ethics*, Wiley, Hoboken, Nj, pp. 142 ff.

Overall we think that the ambiguity surrounding the concept of control is mainly due to a misleading conflation between *conceptualization* and *management* of privacy. In order to reach a refine understanding of the notion of control and its link to privacy theories and polices, it is important to draw a distinction between the different uses and roles of this notion[35]. Accordingly, it is important to differentiate the conceptual dimension of control – i. e. control as conceptual foundation of privacy - from its instrumental dimension – i. e. control as a tool for the management of privacy. In the next section we will focus on the instrumental dimension of control as it has been deployed by EU institutions.

## II. The notion of control in the EU policy documents

Over the last decades, the idea that individuals should have an effective control over their own data has become a key notion of the rhetoric deployed by EU institutions in the field of data protection. In many policy documents, control is advocated as an important tool for protecting privacy and achieving the empowerment of data subjects. In this section, we will explore some of these documents and try to identify the main characteristics of the notion of control as it is featured in EU documents.

Before starting the analysis of these characteristics, it is worth formulating a few preliminary remarks. First, the notion of control is mentioned in documents of very diverse nature, ranging from preparatory works for legislation and legislative text, to experts opinions, to vulgarized material addressed to citizens, etc. This diversity illustrates the pervasiveness of the rhetoric of control in the field of data protection. Second, as we will see hereafter, in most of these documents, the notion of control takes the shape of a toolbox at disposal of the data subjects: they get equipped with a set of subjective "micro-rights" which supposedly enable them to be in control at the different stages of the processing of their data. Third, the notion of control appears to have a much more extended meaning than in the scholarly literature given the operational/instrumental dimension of the EU policy documents[36].

Here is a limited list of EU documents mentioning the notion of control in the recent years:
- 2013 Proposal for a general data protection regulation[37];

---

[32] R. Gavison (1980), Privacy and the Limits of Law. In: *Yale Law Journal*, Vol. 89, pp. 421-472.

[33] Altman, I. (1976), Privacy: A Conceptual Analysis. In: *Environment and Behavior*, Vol. 8, No. 1, pp. 7-29. For Altman, privacy is conceived as a "boundary control process", the selective control over access to oneself.

[34] C. Goodwin (1991), Privacy: Recognition of a Consumer Right. In: *Journal of Public Policy & Marketing*, Vol. 10, No. 1, pp. 149-166. For Goodwin, privacy includes two dimensions of control: control over presence of others in the consumer's environment and control over information dissemination.

[35] Along the same lines, see H. T. Tavani & J. H. Moor (2001), Privacy Protection, Control of Information, and Privacy-Enhancing technologies. In: *Computers & Society*, pp. 6-11. For the authors, any relevant theory of privacy should distinguish between three components: concept, justification and management.

[36] Th. C. Grey (1989), Holmes and Legal Pragmatism. In: *Stan. L. Rev.*, Vol. 41, pp. 805-806.

[37] See Draft of the Regulation of the European Parliament and of the Council on the protection of individuals with regard to the processing of personal data and on the free movement of such data (*General Data Protection Regulation*), Unofficial consolidated version after LIBE Committee vote provide by the Rapporteur, October 22th, 2013.

- 2012 Communication from the Commission, Safeguarding Privacy in a Connected World. A European Data Protection Framework for the 21th Century[38];
- 2012 European Commission brochure and movie "Take control of your personal data"[39];
- 2011 Article 29 Data Protection Working Party, Opinion 15/2011 on the definition of consent[40];
- 2010 Communication from the Commission, A comprehensive approach on personal data protection in the European Union[41];
- 2009 Article 29 Data Protection Working Party, "The Future of Privacy"[42].

For the sake of clarity, we will especially focus on two documents which, despite their distinctive nature, are highly representative of the current rhetoric of control fostered by EU institutions.

The first document is a communication adopted by the European Commission in 2012, titled "Safeguarding Privacy in a Connected World. A European Data Protection Framework for the 21st Century". This communication is part of the works pertaining to the reform of the EU's data protection framework. In the very beginning of this document, it is explicitly stated that "[i]n this new digital environment, individuals have the right to enjoy *effective control* over their personal information."[43] Following that logic, the first half of the communication is entirely dedicated to the thematic of control as it is featured in section 2 "Putting individuals in control of their personal data".

Firstly, the document mentions different issues raised by digital environments which undermine the effectiveness of data protection rules: the lack of harmonization of the member states legislations and the data protection national authorities, the ever-increasing volume of collected data, the perception of loss of control by citizens, etc. Relying on this overview of the situation, the communication then recalls one of the main ambitions of the new legislative act proposed by the Commission: "The aim […] is to strengthen rights, to give people efficient and operational means to make sure they are fully informed about what happens to their per-sonal data and to enable them to exercise their rights more effectively."[44] In order to achieve this aim, the Commission proposes a set of new rules which will: "improve individuals' ability to control their data"; "improve the means for individuals to exercise their rights", "reinforce data security", and "enhance the accountability of those processing data".

---

[38] Communication from the Commission to the European Parliament, the Council, the Economic and Social Committee and the Committees of the Regions, *Safeguarding Privacy in a Connected World. A European Data Protection Framework for the 21th Century*, COM(2012) 9 final, Brussels, 25.01.2012.

[39] European Commission - Directorate-General for Justice, *"Take control of your personal data"*, Luxembourg: Publications Office of the European Union, 2012, http://ec.europa.eu/justice/data-protection/document/review 2012/brochure/dp_brochure_en.pdf.

[40] Article 29 Data Protection Working Party, *Opinion 15/2011 on the definition of consent*, 01197/11/EN WP187, Adopted on 13 July 2011, http://ec.europa.eu/justice/data-protection/article-29/documentation/opinion-recommendation/files/2011/wp187_en.pdf. See pt. II.3. "related concepts"; control is explicitly mentioned among other important concepts relating to consent (alongside with informational self-determination, autonomy and transparency).

[41] Communication from the Commission to the European Parliament, the Council, the Economic and Social Committee and the Committees of the Regions, *A comprehensive approach on personal data protection in the European Union*, COM(2010) 609 final, Brussels, 4.11.2010. See § 2.1.3 titled "Enhancing control over one's own data".

[42] Article 29 Data Protection Working Party, *The Future of Privacy*, Joint contribution to the Consultation of the European Commission on the legal framework for the fundamental right to protection of personal data, Adopted on 01 December 2009, 02356/09/EN, WP 168, http://ec.europa.eu/justice/policies/privacy/docs/wpdocs/2009/wp168_en.pdf. See § 59, p. 15 the reference to the "empowerment of the data subject".

[43] See § 2 (emphasis added).

More precisely, four objectives are mentioned which are supposed to empower the data subjects to "improve individuals' ability to control their data"[45]. Foremost among these objectives, is the principle of consent and the reinforcement of the related legal requirements. This comes as no surprise since consent remains a cornerstone of the EU approach to data protection. Indeed, from a fundamental rights perspective, it is conventionally considered as the "best way for individuals to control data processing activities"[46]. Although consent plays a key role in giving control to data subjects[47], it is not the only way to do this. The Commission also aims at equipping individuals with a right to be forgotten, guaranteeing data accessibility and a right to data portability, and reinforcing the right to information.

In addition to this "bundle" of rights granted to the data subject, the communication equally points to other rules which are deemed to foster a more effective management of personal information. Very interestingly, the Commission seeks to reinforce control by additional rules which are of a radically different nature than the micro-rights granted to the data subject. They consist of an heterogeneous set of organizational and technological tools such as: (i) improved *administrative and judicial remedies*, by strengthening national data protection authorities' independence and powers and enhancing administrative and judicial remedies when rights are violated; (ii) reinforced *security measures*, by encouraging the use of privacy-enhancing technologies, privacy-friendly default settings and privacy certification schemes; (iii) increased *responsibility and accountability*, in particular by requiring data controllers to designate a data protection officer; introducing the "privacy by design" principle and introducing the obligation to carry out data protection impact assessments for organizations involved in risky processing.

At the other end of the spectrum, it is also worth paying attention to a second document issued by the European Commission in 2012. This document explicitly titled "Take control of your personal data" is a small brochure published to raise awareness among EU citizens about the new legal reform and, more precisely, about the changes that will strengthen citizens' rights in the field of data protection[48]. With the help of simplistic slogans, garish fluorescent fonts and fancy drawings, the Commission tries to convey its message to the general public: "Every time you go online you share information about yourself. And the more you do online the more important it is that you and your personal data are protected. The EU is proposing changes that will strengthen your protection online. The new EU laws are designed to *put you in control of your own information* and safeguard your right to personal data protection."[49]

---

[44] See, p. 5.
[45] See, p. 6.
[46] See Committee on Civil Liberties, Justice and Home Affairs (Rapporteur J. Ph. Albrecht), Report on the proposal for a regulation of the European Parliament and of the Council on the protection of individuals with regard to the processing of personal data and on the free movement of such data (General Data Protection Regulation) (COM(2012)0011 – C7-0025/2012 – 2012/0011(COD)), November 21th, 2013, p. 200. See also Opinion 15/2011 of the Article 29 Data Protection Working Party, *op. cit.*, p. 8.
[47] See D. J. Solove, « Privacy Self-Management and the Consent Paradox », *loc. cit.*, pp. 1880-1903; Ch. Lazaro & D. Le Métayer (forthcoming in Autumn 2014), Le consentement au traitement des données à caractère personnel : une perspective comparative sur l'autonomie du sujet. In : *Rev. Juridique Themis*, Vol. 48, No. 3.
[48] The publication of this brochure was accompanied by the release of a short film available online at http://ec.europa.eu/justice/data-protection/minisite/users.html.

Despite its naïve comic strip-like format, it is interesting to observe that the approach to control deployed in this document is structured along the same lines than the aforementioned communication. Indeed, alongside micro rights granted to data subjects[50], the brochure makes also reference to other organizational or technological instruments which are supposed to foster control such as: contact points (where to go and who to talk to in case of problems); privacy by default (make to settings of all websites privacy-friendly; possibility to change the privacy settings yourself); notification breach to the concerned person and the Data protection authority in case the data have been lost or stolen; and more globally the harmonization (same protection everywhere in the EU; high level of protection regardless of where your personal data are stored and handled).

The overview of these two radically different documents published by the EU institutions in the recent years reveals a much more entangled approach to control than it is conceived in scholarly literature. In these documents, the framing of the privacy issue and the implementation of data protection through control has a dual nature, both individual/subjective and organizational/structural.

On one side, the granting of a set of micro-rights to the data subject echoes the main tenets of the control theories: the empowerment of the subject through individual choice and participative agency[51]. In that sense, the concept of control refers to individual's ability to make decision about one's personal data trough autonomous choice. According to the right-based approach embedded in EU policy documents and legislation, the rational and autonomous data subject gets then equipped with tools which will improve the way he or she can control the conditions under which one's personal information is collected, used, and transferred. Eventually, this "legal equipment" aims at transforming the data subject in an active agent who can (and ought to) shape one's own digital live[52].

On the other side, although the individual remains the main agent of control and of the decision-making process in the rhetoric developed by the EU institutions, the notion of control gets somewhat extended beyond strictly individualistic approaches. In particular, the regulator mobilizes a more operational notion of control that cannot be reduced to the purely subjective and mental activity of an autonomous data subject. For the regulator, it is also clear that individual control cannot be exercised without putting a "control architecture" in place[53],

---

[49] The same "slogan" is repeated at the very end of the brochure: "The new EU laws will *put you in control* of the information about yourself that you share online. They'll give you the right to know who's using your data and why and to know if the security of your data is at risk. [...]." (emphasis added).

[50] The brochure especially mentions the right to a better information in order to help people in the decision-making process (of sharing their personal information), the consent requirements (explicit permission and withdrawal of consent), the right to be forgotten (delete permanently the data after one has shared them) and the right to data portability (remove the data and give them to another service provider).

[51] Ch. Fuchs (2011), *op. cit.*, p. 220. In that sense, as Fuchs points out, "control theories are subjective theories [...] because they stress the dependence of privacy on human subjectivity and individual action and choosing."

[52] However, let's note that the regulator imposes in some circumstances substantial limits to its liberty to alienate data (see the case of "sensitive data").

[53] In that regard, we believe that it would be an overtly simplistic critique to assert that EU institutions simply reproduce or contribute to the reinforcement of the neo-liberal ideology of empowerment of the self, which encourages people to see themselves as their self-entrepreneurs and to behave as active subjects responsible for

The control over personal data: True remedy or fairy tale ?namely a set of structural measures that aim at creating a reliable and secure environment for the data subject. Framed in such a way, control over personal data is not treated solely as a matter of individual negotiation and party autonomy in contracting arrangements, but as an operation that is also potentially dependent on other important "environmental" variables: technological (i.e. security measures, privacy by default settings, etc.) and organizational (i.e. accountability of data controllers, privacy impact assessment, etc.).

On closer examination, the analysis of the EU documents shows the diversity of normative tools that the data subject has to be equipped with in order to get control over one's own data and, more globally, to keep up with technological change[54].

### III. Control from the technical point of view

The previous sections have shown that the notion of control is multifaceted, but most of the interpretations of control, if not all, assume that the subject must be able to act, in one way or another, to exercise his rights. In the digital world, these actions are mostly carried out through information technologies. A relevant question at this stage is therefore: what does control mean in the technical world and can technologies provide appropriate tools to support the notions of control proposed by lawyers and philosophers?

First, it is worth noting that the view of control as a set of "micro-rights" in the fundamental rights perspective is very much in line with the view of control in computer science[55]. One of the most common uses of the term "control" in this area can be found in the expression "access control" in computer security. Access control can be seen as a technique for restricting access to a resource (for example a personal data) to authorised users. Interestingly, a difference is made in computer security between discretionary access control, in which the owner of an object defines the rules, and mandatory access control, in which the rules are defined by the system administrator. This difference raises the question of who is the actor in charge of deciding, or who is "in control".

Another interesting observation about the notion of control in computer security is that its use has been extended to "usage control", precisely to provide ways to implement legal provisions in the area of intellectual property. In contrast with access control, usage control makes it possible to control the object also during its usage, for example to enforce time limitations or a maximum number of uses. Usage rights can also be conditioned to certain obligations. Usage control can typically be useful to implement Digital Right Management, but it has also been

---

enhancing their own well-being. See N. Rose (1992), Governing the enterprising self, In: P. Heelas & P. Morris (eds.), *The values of the enterprise culture. The Moral Debate*, Routledge, London – New York, pp. 141-163. In the field of data protection, the overall picture seems to us far more complex.

[54] Such a diversity of tools is also characteristic of the "mixed approach" developed in the new Proposal for a general data protection regulation. Although the data subject's control is one of the strategic objectives targeted by the Proposal, there is no explicit reference to the concept of control among the legal provisions. However, it is mentioned in recital 6 which very basically states that "*individuals should have control of their own personal data*", while insisting on the need to build strong and more coherent data protection framework in the Union in order to foster the digital economy and to reinforce certainty for all the actors (emphasis added). See also recital (51a).

[55] The word control » has been used in different areas of computer science, but we focus here on the uses of the word having a connection with privacy or exhibiting features which can be transposed to privacy.

suggested to apply it to personal data management, for example to limit the use to the declared purpose.

In the context of operating systems, control also includes other "micro-rights" such as the rights to create, read, modify or delete a file, or to get access to a directory list and these rights can be granted to individuals or to groups of users.

To sum up, the different variants of control in computer science can be classified according to two main criteria:
- What is the subject of the control (who is supposed to be in control)?
- What is the object of the control (what does the subject control)?

As far as privacy enhancing technologies are concerned, one must admit that they mostly reflect the individualistic view discussed in Section 2. However, as we will show hereafter, the collective dimension of control is also supported by some recent tools. With respect to the object of the control, three main categories of tools can be identified: the first one, sometimes called TETs (for Transparency Enhancing Technologies), basically supports the right to be informed, the right for the subject to know what happens to his personal data; the second one, supports all "active rights" of the subject such as the right to express his consent, or to have his data modified or deleted; the third one supports "negative rights", such as the right to prevent the disclosure of his data (or to ensure the implementation of the "data minimization" principle).

In the remainder of this Section, we first study the object of control (and the aforementioned three types of rights) in Subsection 3.1 before discussing the subject of control (and the individual versus collective views) in Subsection 3.2 and concluding with some reflexions on the relativity of control in Subsection 3.3.

**A. The object of control: a set of micro-rights**

The exercise of control rights of the subject requires a deliberate action on his part, which means not only that the system should make this action possible but also that it should provide sufficient information to the subject to ensure that he can properly understand the situation and take well-informed decisions. The first type of technology that provides valuable support to the subject in this phase is sometimes called the "transparency enhancing technologies" (TET)[56].

*TETs (for Transparency Enhancing Technologies): the right to be informed*

TETs can take different forms depending on the context and the type of information provided to the user. As far as web sites are concerned, the simplest forms of TET are the "privacy icons" which are visual signs designed to make it possible to get at a glance the main features of the privacy policy of the site (data collected, purpose, deletion delay, etc.). Users can then parameterize their privacy policy in such a way that their browser can automatically check that the policy declared by a site meets their own policy and to inform them (for example through specific icons) of the result of the verification[57].

---

[56] Mireille Hildebrandt, Bert-Jaap Koops. "The Challenges of Ambient Law and Legal Protection in the Profiling Era" The Modern Law Review 73.3 (2010): 428-460.

Some websites also provide a dashboard functionality informing users about the personal data stored[58] and what third parties can get access to it. But this kind of site must have very carefully designed interfaces to ensure that they do not mislead users[59]. For example, the European PrimeLife project has proposed a Firefox extension called Privacy Dashboard, that allows users to know some of the practices of the websites they are using, for example whether they use cookies, geolocation, third party content or other tracking means. An icon displays a more or less happy face depending on the overall evaluation of the web site[60].

Specific solutions have also been proposed to improve the privacy interfaces of social networks, for example to reduce unnoticed over-sharing of information, to make it easier to find out to whom a particular attribute is visible[61] or to help users avoiding to make posts that they may later regret[62].

Personal data are sometimes collected without the subject being aware of it and by parties that he has never heard about. This happens typically through cookies created on his computer while he is browsing and used by a variety of companies to track his activities and ultimately to serve him personalized advertisements based on his browsing profile. Users can get a picture of the tracking going on behind their back using a tool like Lightbeam[63] (formerly Collusion) which is a Firefox add-on recording the events associated with the visited sites and allowing users to display a graph showing the tracking sites and their interactions. Several tools such as TaintDroid[64] or Mobilitics[65] have also been proposed for smartphones which represent another major source of leak of personal data.

*Active rights: consent, modification, deletion, etc.*

When he has reached a decision about his privacy preferences, the next step for the data subject is to express this decision. Several techniques are available to help him in this task. They differ mostly in terms of scope (general purpose versus specific) and interfaces. On the general purpose side, a number of languages have been proposed to express privacy policies[66]. The general principle is that both the subject and the controller (typically a web site) should be able to author privacy policies which are translated into a machine readable format. The policies can then be processed automatically and matched to ensure that a controller collects only personal data associated with a privacy policy (defined by the subject) consistent with his own policy. As an illustration, tools like P3P[67] and Privacy Bird[68] allow respectively websites to declare their privacy policies and visiting users to have these policies analyzed and compared with their own preferences. Depending on the result of the matching, different icons can be displayed in order to inform the user and let him either accept to visit the site, refuse, or look further into his privacy policy (in which case, Privacy Bird can also be used to display the policy in a user-friendly way, starting with a summary). The preferences of the user can be set through a number of panels allowing him to choose different levels of protection for different types of data (health, financial, etc.).

---

[57] Privacy Bird is an example of browser add-on (for Internet Explorer) that provides this facility
[58] Google Dashboard provides this facility but it shows only a subset of the collected data.
[59] For example Scott Lederer et. al. identify five pitfalls for designers (obscuring potential information flow, obscuring actual information flow, emphasizing configuration over action, lacking coarse-grained control and inhibiting existing practice) and they show existing systems falling into these pitfalls or avoiding them.
[60] http://primelife.ercim.eu.
[61] Thomas Paul, Daniel Puscher, Thorsten Strufe: Improving the Usability of Privacy Settings in Facebook. CoRR abs/1109.6046 (2011).
[62] Yang Wang, Saranga Komanduri, Pedro Leon, Gregory Norcie, Alessandro Acquisti, Lorrie Faith Cranor. "I regretted the minute I pressed share": A Qualitative Study of Regrets on Facebook. Proceedings of the Seventh Symposium on Usable Privacy and Security (SOUPS), ACM, July, 2011.
[63] http://www.mozilla.org/en-US/lightbeam.
[64] Enck, William and Gilbert, Peter and Chun, Byung-Gon and Cox, Landon P. and Jung, Jaeyeon and McDaniel, Patrick and Sheth, Anmol N. TaintDroid: An Information-flow Tracking System for Realtime Privacy Monitoring on Smartphones, Proceedings of the 9th USENIX Conference on Operating Systems Design and Implementation, OSDI'10, 2010, pp. 1-6.
[65] Mobilitics: analyzing privacy leaks in smart phones, Jagdish Prasad Achara, Franck Baudot, Claude Castelluccia, Geoffrey Delcroix and Vincent Roca, ERCIM News, 93, April 2013.

However, this approach raises several challenges. For this kind of consent to be legitimate from a legal point of view, it must be free, specific, informed and unambiguous[69]. For example, the categories of data that can be used in P3P or Privacy Bird may be too coarse in many situations and force the users to disclose more data or grant to third parties broader rights than they would really like to. Most languages may also lead to ambiguities or statements that can be interpreted in different ways. Ambiguities may come from the use of vague terms but also from the combination of rules or the default rules[70]. One of the criticisms raised against early privacy frameworks such as P3P was their lack of clarity and the divergent (or even misleading) representations of privacy policies[71]. An option to solve the ambiguity problem is to re- sort to a sound, mathematical definition of the semantics of the language. This approach has been followed in several proposals, such as CI[72] and S4P[73] and SIMPL[74]. Even though their scope goes beyond the definition of privacy policies[75] and they may have a strong impact in the future, these academic results have not moved out into the field yet. One reason why these languages have not been deployed yet is the fact that their generality raises new challenges in terms of user interface: to be really usable, they should be integrated within tools allowing users to express their choices in a convenient and efficient way.

---

[66] See for example : Adam Barth, Anupam Datta, John C. Mitchell, Helen Nissenbaum. Privacy and Contextual Integrity: Framework and Applications, Proceedings of the 2006 IEEE Symposium on Security and Privacy, SP '06, IEEE Computer Society, pp. 184-198, 2006. Moritz Y. Becker, Alexander Malkis, Laurent Bussard. S4P: A Generic Language for Specifying Privacy Preferences and Policies, Technical report MSR-TR-2010-32, Microsoft Research . D. Le Métayer. A formal privacy management framework, Proceedings of the FAST'2008 Workshop (IFIP WG 1.7 Workshop on Formal Aspects in Security and Trust), Springer Verlag, Lecture Notes in Computer Science. A. Barth, J. C. Mitchell, A. Datta, S. Sundaram. Privacy and utility in business processes. Proc. CSF, pp. 279-294, 2007. G. Karjoth, M. Schunter, E. V. Herreweghen. Translating privacy practices into privacy promises, how to promise what you can keep. Proc. POLICY, pp. 135-146, 2003.
[67] W3C. Platform for privacy preferences (P3P). W3C recommendation. www.w3.org. Technical report, W3C, 2002.
[68] http://www.privacybird.org.
[69] European Parliament and the Council of the European Union. Directive 95/46/EC of the European Parliament and of the Council of 24 October 1995 on the Protection of Individuals with Regard to the Processing of Personal Data and on the Free Movement of such Data. Brussels: European Parliament, 1995.
[70] C. A. Brodie, C.-M. Karat, and J. Karat. An empirical study of natural language parsing of privacy policy rules using the Sparcle policy workbench. In Symposium On Usable Privacy and Security (SOUPS), 2006.
[71] Reidenberg, Joel and Cranor, Lorrie Faith, Can User Agents Accurately Represent Privacy Policies? (August 30, 2002). Available at SSRN: http://ssrn.com/abstract=328860.
[72] Adam Barth, Anupam Datta, John C. Mitchell, Helen Nissenbaum. Privacy and Contextual Integrity: Framework and Applications, Proceedings of the 2006 IEEE Symposium on Security and Privacy, SP '06, IEEE Computer Society, pp. 184-198, 2006.
[73] Moritz Y. Becker, Alexander Malkis, Laurent Bussard. S4P: A Generic Language for Specifying Privacy Preferences and Policies, Technical report MSR-TR-2010-32, Microsoft Research.
[74] D. Le Métayer and S. Monteleone. Computer assisted consent for personal data processing. 3d LSPI Conference on Legal, Security and Privacy Issues in IT, 2008.

Another option provided by most browsers is the Do Not Track[76] feature that allows users to express a choice not to be tracked in their browsing activities. This opt-out choice is communicated to visited websites through a specific DNT HTTP header sent every time data is re- quested from the Web. However, there is no consensus yet on how web sites should precisely interpret this DNT signal. As a result, many sites simply ignore them while others may limit the amount information that they collect.

More generally, the actual enforcement of the privacy choices of the data subject depends very much on the localization of the personal data. In fact, the only decisions of the data subject that can be enforced locally are his choices concerning cookies, popups or ad blockers. These choices are implemented on the device of the subject as browser (or Flash Player) options or extensions. As long as they know how to do it, subjects can also decide at any time to erase cookies stored on their computer[77] or their browsing history. The enforcement of all other types of consent rely on the existence of appropriate technical means on the side of the controller (and, in some cases, of other stakeholders) and their proper execution. The subject has therefore no choice but to put some trust on the data collector: he must trust him for having such technical means in place (and not trying to bypass them). As discussed in the next subsection, this trust could be enhanced through compliance audits conducted by independent third parties.

As suggested above, one technical option to ensure the implementation of privacy policies is to resort to DRM like technologies to monitor the use of personal data[78]. Personal data would then be managed in the same way as digital content (e.g. video or music) but it is not clear whether this solution is really viable considering that personal data would easily be copied after it is output from the DRM system to be used for the purpose. Experience has also shown that DRM techniques can often be bypassed with moderate effort. As stated by Manuel Hilty, David Basin and Alexander Pretschner, "at the very least, DRM can act as a support mechanism … and thereby increase the likelihood that the obligations are fulfilled, or at least pre- vent unintended violations resulting from carelessness."[79] Another extreme solution would be to require data controllers to use a trusted computing environment to process personal data. Such a trusted platform ensures that the system behaves as expected at the price of having a unique encryption key loaded in the hardware and made inaccessible to the user.

---

[75] They can be used to specify norms, in a more general sense, for example CI has been applied to HIPPA (Health Insurance Portability and Accountability Act), COPPA (Children's Online Privacy Protection Act) and GLBA (Gramm-Leach-Bliley Act).
[76] https://www.mozilla.org/en-US/dnt.
[77] This may not always be obvious for non-technical users though, for example they may not be aware of the fact that different types of cookies may be stored on their computer, some of them directly by their browser, others by Adobe Flash Player, which require different actions.
[78] Mayer-Schönberger, V. (2006) Beyond Copyright: Managing Information Rights with DRM. 84 Denver University Law Review 181
[79] Hilty, M., Basin D., Pretschner, A.: On obligations, Proc. 10th European Symp. on Research in Computer Security (ESORICS'05), Springer LNCS 3679, pp. 98-117, 2005.

This solution has been used in specific cases such as healthcare information processing[80] but it remains to be seen whether it can become a more widely adopted solution considering the controversies about the trusted computing technology itself, which results in a loss of control of the users on their own computers[81] (and an increased control of the computer manufacturers and software providers).

*Negative rights: non-disclosure, data minimization*

Many other technologies (sometimes called "privacy enhancing technologies" or PETs) are available to enforce privacy rights[82]. The main goal of PETs is to reduce as much as possible (or even to prevent) the disclosure of personal data (typically through the use of cryptographic techniques). For example, it is possible to use PETs to implement a smart metering system in which the operator does not get any personal data of the users (apart from their quarterly fee). This is made possible through a combination of architectural choices (the fee is computed locally, on the equipment of the users) and appropriate cryptographic protocols (to ensure that the users are accountable: they cannot cheat on the computation of the fee).

This notion of "privacy by architecture" differs from the usual vision of "privacy by control" since the user does not have to take any action: the design of the system ensures that his or her personal data will not be disclosed. If to use the Latourian terminology, one can say in that case that control is entirely delegated to non-human actors.

This analysis is also in line with the distinction made by some authors between hard privacy and soft privacy[83], which are associated with different trust assumptions: hard privacy (as illustrated by PETs) tries to avoid as much as possible placing any trust in any third party (or to reduce this trust), while soft privacy is based on the assumption that the subject will, technically speaking, lose control over his data and therefore will have no choice but to place a certain amount of trust in the data controller. In the latter situation, technologies for enforcing the rights of the subject can then be seen as ways to reduce this "loss of control" and to organize it in the best interest of the subject (information, consent delivery, accountability, etc.).

---

[80] http://www.trustedcomputinggroup.org/files/resource_files/3B1360F8-1D09-3519-AD75FFC52338902D/03-000216.1.03_CBIHealth.pdf

[81] Which is, admittedly, the intended effect for privacy enforcement when the trusted execution environment is on the side of the data controller, since the objective is to force him to fulfill the sticky privacy policies.

[82] Privacy-Enhancing Technologies for the Internet. Ian Goldberg, David Wagner, Eric A. Brewer, IEEE COMPCON '97, February 1997. Privacy-Enhancing Technologies for the Internet III: Ten years later. Ian Goldberg, Chapter 1 of "Digital Privacy: Theory, Technologies, and Practices", Alessandro Acquisti, Stefanos Gritzalis, Costos Lambrinoudakis, Sabrina di Vimercati, editors, December 2007. A critical review of 10 years of privacy technology. George Danezis and Seda Gürses. Surveillance Cultures: A Global Surveillance Society?, April 2010, UK. Understanding the landscape of privacy technologies. Claudia Diaz and Seda Gürses. Extended abstract of invited talk in proceedings of the Information Security Summit, pp. 58-63, 2012. PETs in the surveillance society: a critical review of the potentials and limitations of the privacy as confidentiality paradigm. Seda Gürses and Bettina Berendt. In: Serge Gutwirth, Yves Poullet, Paul De Hert (Eds.): CPDP 2009, Springer Verlag. Privacy Enhancing Technologies: A Review. Yun Shen, Siani Pearson. HP Laboratories HPL-2011-113.

[83] Mina Deng, Kim Wuyts, Riccardo Scandariato, Bart Preneel, Wouter Joosen, A privacy threat analysis framework: supporting the elicitation and fulfillment of privacy requirements, Requirements Engineering, Springer, Volume 16, Issue 1, Pages 3-32, Special Issue on Digital Privacy, March 2011.

## B. The subject of control: individual versus collective views

The above discussion about "privacy by architecture" versus "privacy by control" also echoes the debate on the "individualistic" versus "collective" views of control and privacy: "privacy by architecture" can be seen as a form of collective control because the decision to implement privacy protections is imposed by some authority which is supposed to represent the interests of the subjects as a whole. Ideally, the design of the system could even be approved or certified by an independent third party.

This collective view of privacy, even if not dominant in the technological landscape, is supported by other types of tools. For example, regardless of the actual level of information that they can obtain, one could argue that individuals are always in a weak position when they have to take decisions about the disclosure of their personal data because they generally do not have the necessary expertise to fully understand all legal and technical aspects of the situation. One solution to redress this imbalance is to provide some form of collaboration between individuals to help them analyze privacy policies and warn their pairs about inacceptable terms. ToS;DR[84] (Terms of Service; Didn't Read) is an example of effort in this direction. The goal of ToS;DR is to create a database of analyses of the fairness of privacy policies and to make this information available in the form of explicit icons (general evaluation plus good and bad points) which can be expanded if needed into more detailed explanations. Users can also install a browser add-on to get the ratings directly when they visit a page. A key aspect of ToS;DR is the fact that users can submit their own analysis for consideration, the goal being that, just like Wikipedia, a group consensus will emerge to provide a reliable assessment of each policy. This type of tool is especially interesting as they promote a broader notion of control, less individualistic and more collective (even if the final decision always pertains to the subject).

Accountability (at least in its strongest forms, when it requires auditability by independent third parties) can also be seen as a form of collective approach in the sense that it ensures that the subject is not left alone in front of the data controller. The key idea behind the notion of accountability is that data controllers should not merely comply with data protection rules but should also be able to demonstrate compliance or "showing how responsibility is exercised and making this verifiable", as stated by the Article 29 Working Group[85]. Technologies can facilitate accountability through the privacy policy languages and frameworks mentioned above. They can also contribute to accountability of practices to ensure that data controllers can be in a position to demonstrate that their practices, hence their use of personal data, complies with their obligations. The main piece of evidence for accountability of practices should be execution logs recording all events that can have a potential impact on this compliance. Technologies are available to support log secure storage[86] and their use to conduct audits[87].

---

[84] http://tosdr.org.

[85] Article 29 Data Protection Working Party. Opinion 3/2010 on the principle of accountability. 2010. http://ec.europa.eu/justice/policies/privacy/docs/wpdocs/2010/wp173_en.pdf.

[86] M. Bellare and B. S. Yee. Forward integrity for secure audit logs. Technical report, University of California at San Diego, 1997. B. Schneier and J. Kelsey. Secure Audit Logs to Support Computer Forensics.

[87] See for example: D. Garg, L. Jia, A. Datta. Policy Auditing over Incomplete Logs: Theory, Implementation and Applications, in Proceedings of 18th ACM Conference on Computer and Communications Security, October 2011. Log Analysis for Data Protection Accountability — Denis Butin and Daniel Le Métayer — 19th International Symposium on Formal Methods (FM 2014), Springer Verlag LNCS Volume 8442, 2014, pp 163-178.

## C. Control as a relative notion

Paradoxically, the term "control" as interpreted by lawyers seems to be used as a key privacy principle in situations where "control", in the technical sense, is effectively relinquished, or at least shared, by the subject. Indeed, in most situations, subjects actually lose the control over their personal data as soon as they disclose them in the sense that they cannot have 100% guarantees concerning the use of their data by the data controller. This should not imply that control is a meaningless or illusory principle though, but this observation argues in favour of an interpretation of control as a relative notion. The main lesson to be drawn from this analysis is therefore that technical means are available to enhance control but lawyers and policy makers should avoid overreliance on this notion of control because it cannot be, from a practical point of view, an absolute protection.

## IV Conclusion

More than ever the notion of control plays a pivotal and pervasive role in the discourse of privacy and data protection. Privacy scholarship and regulators propose to increase individual control over personal information as an ultimate prescriptive solution: it is considered as a crucial means of management of digital identity and empowerment of the data subject. Nevertheless, at a time of ever-increasing digitalization of life and global circulation of data, such rhetoric seems at odds, if not totally paradoxical. Indeed the premise of autonomy and active agency implied in this rhetoric seems to be radically undermined in the context of contemporary digital environments and practices. Exploring this ambiguity from an interdisciplinary perspective, this report passes in review the different meanings of the notion of control and tries to clarify the characteristics of this notion as it is developed in several sources of the literature and EU policy documents.

As we have seen, the policy or regulatory initiatives in the field of data protection described in this report represent a more entangled approach to control than the strict individualistic paradigm of the "privacy as control" theory developed in the scholarly literature. In the EU policy documents, control is conceived as a dual notion, both individual and structural. In the eyes of the regulator, the burden of controlling personal information cannot only weigh on the data subject 'shoulders. For control and the related micro-rights granted to the data subject to be effective, it has to be supported by structural measures.

The (ab)use of the fashionable rhetoric of control by policymakers tends then to obscure this structural dimension, but even a cursory review of the EU policy documents reveals that the idea of control is not dissociated from the implementation of organizational and technical measures. This shows that the regulator is aware that control over personal information cannot only be a matter of individual agency. To be sure, control cannot be properly achieved if the data subject is not put in a position to monitor that the data controller actually complied with his privacy preferences. Similarly, control becomes almost impossible when the data subject has to deal with privacy-unfriendly default settings and technologies. Therefore the regulator seeks to reinforce individual control by additional rules which consist of a heterogeneous set of organizational and technological tools that foster, for instance accountability and privacy by design mechanisms.

Despite their appeal to a much more extended and operational meaning of the notion of control, we would like to argue that EU policymakers fall short of grasping the crucial issues raised by the notion of control. This is mainly due to the fact that they still remain excessively attached to the individualistic paradigm according to which the data subject is depicted on the basis of the conventional figure of the "rational and autonomous agent", this monadic and abstract individual capable of deliberating about personal goals and of acting under the direction of such deliberation[88]. The reliance on this overtly simplistic account of human agency impedes to further investigate the pragmatic modalities of the operation of control and, more specifically, to apprehend the normative consequences of two fundamental questions: control of *what* and control *by whom* and? Taking control seriously requires then raising the issues of the *object* and the *subject* of control.

The first question raised by the thematic of control relates to the definition of its object. What is the target of individual control and can it be limited to personal information as it is defined in data protection legislation? What does it really mean to be in control of one's data in the context of contemporary socio-technical environments and practices? Nowadays individual control can certainly not be considered as a panacea to solve the thorny issues raised by "Big Data phenomenon" and the ever-evolving data mining and profiling practices. In particular, individuals are often not aware or do not understand how this happens, who collects their data, nor how to exercise control because the use and further transfer of personal data is very often done in an extremely complex and non-transparent way. This situation of course strongly undermines the very idea of control.

Besides the voracious collection and use of personal information, the big data phenomenon also raises the issue of control over large amounts of data which cannot be included in the category of "personal data" as it is currently defined by the legislation. Indeed, the construction of profiles by private and public organizations is based as much (if not more) on these "impersonal" data as on personal data voluntary (or not) provided by the individuals. For some scholars, one of the main drawbacks of the current EU legal framework is its excessive and misleading focus on the so-called personal data. As A. Rouvroy contends, this "fetishisation"[89] of personal data contributes to obscure the real impact of data mining and profiling activities: the capacity of data subjects to exercise control over their "profiles"[90] and the development of an effective right of data management, especially with regard to data qualified as behavioral and inferred from digital practices. Moreover, the conventional opposition be- tween personal and anonymous data tends to fall apart as "reidentification" techniques get more sophisticated, allowing computer scientists to "deanonymize" individuals hidden in anonymized data with disconcerting ease[91]. For C. Priens, "[t]his then will require a debate on the role of the public domain in providing the necessary instruments that will allow us to know and to control how our behavior, interests and social and cultural identities are 'created'."[92]

---

[88] S. R. Peppet (2011), op. cit., p. 1188. The author brilliantly observes that, despite their effort of reconceptualization, privacy scholars did not abandon the idea "that individuals do and should remain the locus of the decision-making about their personal data".

[89] A. Rouvroy (2014), Des données sans personne: le fétichisme de la donnée à caractère personnel à l'épreuve de l'idéologie des Big Data, In : Le numérique et les droits et libertés fondamentaux, Etude annuelle du Conseil d'Etat, La Documentation française, Paris, p. 418. For A. Rouvroy, the current "fetishisation" of personal data, which is reinforced by the characteristics of the legal framework itself, contributes to obscure the real nature of the problems raised by the phenomenon of Big Data. See also B. Schermer (2011), The limits of privacy in automated profiling and data mining, In: Computer Law & Security Rev., Vol. 27, pp. 45-52.

[90] A. Rouvroy & Th. Berns (2013), op. cit., p. 179. The authors insist on the loss of individuals' control over "their" profiles.

[91] P. Ohm (2010), Broken Promises of Privacy: Responding to the Surprising Failure of Anonymization, In: UCLA L. Rev., Vol. 57, pp. 1701-1778.

The second fundamental question raised by the thematic of control relates to the determination of its subject/agent and implies to interrogate the skills and competences of contemporary data subjects. Who are these data subjects and are they really able to cope with the ever- growing complexity of digital environments? Who can be said to be "in control"?

As it has been unveiled these last years by research in the field of behavioral economics, cognitive sciences or human-computer interaction, the complexity is such that our judgments in this area are prone to errors, stemming from lack of information or computational ability; or from problems of self-control and biased decision-making process. For instance, people time and attention are limited. It is impossible to control every single piece of information about oneself which circulates on the networks through myriads of channels and databases. Moreover, while sharing their data with the members of social networks or with various providers, people might suffer from an "illusion of control"[93]. Another consequence of the emphasis on active choosing/control is the difficulty raised by the situations where people prefer not to choose. Indeed, the costs imposed on data subjects can be so high in complex and technical areas they are unfamiliar with that majority of them tend to "stick" to default options instead of exercising their freedom of choice and being in control of the situation[94]. How can one conciliate the idea of control with the cognitive and behavioral biases that hamper users' privacy and security decision making?

On a more analytical level, let's note that even in the hypothetical case where the data subjects would be perfectly aware and competent, the logics of control assumes perhaps too rap- idly that voluntary disclosure of personal information causes no privacy problems[95]. However, we believe that it is nearly impossible for that data subjects to really measure the breadth of their disclosure and the long term effects of their actions. It is then very unlikely that they can suffer no harm even from a potentially informed, autonomous and responsible decision[96].

---

[92] C. Priens (2006), When personal data, behavior and virtual identities become a commodity: Would a property rights approach matter?, In: Script-ed, Vol. 3, No. 4, p. 272.

[93] Brandimarte, L., Acquisti, A., Loewenstein, G., and Babcock, L., (2009) Privacy Concerns and Information Disclosure: An Illusion of Control Hypothesis, https://www.ideals.illinois.edu/handle/2142/15344. These schol- ars formulate the hypothesis that individuals might suffer from an "illusion of control" when dealing with the publication of their data: "Namely, we hypothesize that when subjects are personally responsible for the publica- tion of private information online, they may also tend to perceive some form of control over the access of that information by others, thereby confounding publication with access."

[94] The issue at stake here is: how can control theory conciliate active choosing/control with "the choice of not to choose"? See C. Sunstein (2014), Active Choosing or Default Rules? The Policymaker's Dilemma, http://papers.ssrn.com/sol3/papers.cfm?abstract_id=2437421. 93

[95] Let's note that one of the logical limits of the control theory becomes obvious in the case of total disclosure. What if a data subject willingly decides to reveal every single piece of information about oneself? Can one still sustain that he retains privacy? One can have control, but no privacy…: "The prospect of someone revealing all of his or her personal information and still somehow retaining personal privacy, merely because he or she retains control over whether to reveal that information, is indeed counter to the way we ordinarily perceive of privacy." See H. T. Tavani (2000), *op. cit.*, p. 2.

[96] Although recent literature has tried to reinvestigate the concept of privacy, there is still much work to do when it comes to identify and evaluate what constitutes privacy harm and its link with (loss of) control over personal data. See Calo, M. R. (2011), The boundaries of privacy harms. *loc. cit.*, pp. 1131-1162 ; D. J. Solove, (2006), A Taxonomy of Privacy. In: U. Penn. L. Rev., Vol. 154, No. 3, pp. 477-560.

For these different reasons, the issue of the agent of control should be addressed with much more caution and attention than it is currently the case. Although proponents of the control theory and policymakers rightly recognize the importance of control, they put so much emphasis on its subjective dimension that they fail to adequately capture the limits of the normative and technical tools put at the disposal of data subjects. If actual empowerment and meaningful autonomy of data subjects are to be achieved, granting them "micro-rights"[97] and providing them with privacy management technologies is certainly not enough. Indeed, the complexity of digital environments and practices is such that one should not expect data subjects to become privacy experts[98] and bear all the risks and responsibilities of privacy management alone. For control to become more than an empty notion, one should therefore definitely embrace the idea that people act and transact in society not simply as individuals in an undifferentiated social world, but as individuals in certain capacities, in distinctive socio-technical contexts. This necessarily implies to integrate in our understanding of the data subject's agency the inescapable collective dimension of control. To put it simply, control over information cannot become effective as long as is not conceived and implemented in terms of shared engagement and cooperation between different human and non-humans actors[99].

On one hand, the various modes of cooperation with non-human actors and the delegation of action to machines has to be tackled more carefully. In digital environments, the exercise of control is highly mediated by technical devices which can enhance but also hinder an agent's capacity to make choices and determine the course of his or her action. In that regard, privacy management technologies deemed to provide more transparency and to allow more granular control over privacy settings do not necessarily solve the users' problems because they can increase their cognitive costs, without addressing the underlying cognitive and behavioral biases[100]. As we have seen in section 3, the diversity of technical tools at the disposal of data subjects as well as their intrinsic working often adds another layer of complexity.

On the other hand, treating control over personal data solely as a matter of individual negotiation and party autonomy in contracting arrangements neglects the underlying relations of powers between actors as well as the collective impact of privacy management which goes beyond individual welfare. In that regard, making control meaningful implies envisioning and creating new modes of relations and cooperation between human actors (data subjects, public institutions, and private organizations) which would enable a much more balanced distribution of risks and responsibilities. In the Proposal for a general data protection regulation, the EU legislator already took a few steps into this direction, by imposing new obligations on data controllers[101] and by taking into consideration the situations where there is a significant imbalance between parties[102]. Alternatives to classical regulation such as "nudge"[103] or "crowdsourcing"[104] could also presumably offer new ways to make control more effective.

---

[97] Such as the right to be informed, the right to revoke consent, the right to access, modify, rectify, and delete the data, the right to data portability, the right to be forgotten, the right to object, etc.

[98] See D. Solove (2013), op. cit., p. 1901: "With the food we eat and the cars we drive, we have much choice in the products we buy, and we trust that these products will fall within certain reasonable parameters of safety. We do not have to become experts on cars or milk, and people do not necessarily want to become experts on privacy either. Sometimes people want to manage their privacy in a particular situation, and they should be able to do so. But globally across all entities that gather data, people will likely find self- management to be a nearly impossible task."

[99] See B. Nardi (1992), Studying context: A comparison of activity theory, situated action models and distributed cognition. In: Proceedings East-West HCI Conference, 4-8 August, St. Petersburg, Russia. pp. 352- 359, http://www2.physics.umd.edu/~redish/788/readings/nardi-ch4.pdf: "[…] it is not possible to fully understand how people learn or work if the unit of study is the unaided individual with no access to other people or to artifacts for accomplishing the task at hand." See also B. Latour (1999), Pandora's Hope: Essays on the Reality of Science Studies, Harvard University Press (chapter 6, titled "A Collective of Humans and Nonhumans: Following Daedalus' Labyrinth").

[100] Moreover, this raises the problem of "familiarity" with technical artifacts and the nature of human agency which is involved in data subjects 'daily idiosyncratic engagements with machines. See L. Thévenot, (2002), *op. cit.*, pp. 53-87.

In offering a brief overview of the two fundamental question raised by the thematic of control, our goal is to foster discussion and encourage a more nuanced understanding of the concept of control. For the empowerment of the data subject to be effective, we believe that there is an urgent need to develop an account of agency of data subjects which takes into consideration the multi-dimensional and varied intersections between individual capabilities and socio-technical environments, including the engagement of the individuals in meaningful participation and collective activity. In the absence of such reconceptualization, the idea of control over personal information pervading current legal and political debates about privacy will amount to nothing more than a fairy tale.

---

[101] See, for instance, the new provisions regarding responsibility and accountability, privacy impact assessment, the notification of a personal data breach to the supervisory authority.

[102] See Article 7, § 4 of the Proposal: "Consent shall not provide a legal basis for the processing, where there is a significant imbalance between the position of the data subject and the controller."

[103] R. Calo (2014), Code, Nudge or Notice?, In: Iowa L. Rev., Vol. 99, pp. 773-803.

[104] See supra, section 3.2.